# On-demand quantum correlation control using coherent photons

Byoung S. Ham

School of Electrical Engineering and Computer Science, Gwangju Institute of Science and Technology
123 Chumdangwagi-ro, Buk-gu, Gwangju 61005, S. Korea
(Submitted on Nov. 22, 2020; bham@gist.ac.kr)

**Abstract:** Over the last several decades, quantum entanglement has been intensively studied for potential applications in quantum information science. Although intensive studies have progressed for nonlocal correlation, fundamental understanding of entanglement itself is still limited. Here, the quantum feature of anticorrelation, the so-called HOM dip, based on probabilistic entangled photon pairs is analyzed for its fundamental physics and compared with a new method of on-demand entangled photon pair generations using coherent light. The fundamental physics why there is no $g^{(1)}$ correlation in HOM dip measurements is answered, and new coherence quantum physics is proposed for macroscopic quantum entanglement generations.

**Introduction**

Since the seminal papers on the Bell inequality [1] and anticorrelation [2] using a beam splitter (BS), quantum entanglement [3-5] has been intensively studied for potential applications in quantum computing [6-10], quantum cryptography [11-15], and quantum sensing [16-20]. For anticorrelation, the so-called a HOM dip [2] has been the proof of an entangled photon pair interacting on a BS via coincidence detection measurements, resulting in $g^{(2)}(0) < 0.5$, where this anticorrelation is a direct proof of the nonclassical feature in terms of maximal correlation between the interacting photon pairs [1-5]. In anticorrelation or the Bell inequality, however, the amplitude correlation $g^{(1)}$ has been completely forgotten according to the particle nature of the photons. Recently, the nonclassical feature $g^{(2)}(\tau) < 0.5$ has been demonstrated even for independent light sources such as sunlight and molecule-generated photons [21,22]. However, the fundamental principle of entanglement generation itself is not still clearly understood.

Based on Heisenberg's uncertainty principle between conjugate variables, e.g., time and frequency, or the photon number and phase, the critical $g^{(2)}$ value of anticorrelation is closely related with the $g^{(1)}$ value of coherence [23]. Under the uncertainty principle, one property is thus coupled with another within a minimum value. In that sense, the wave nature of $g^{(1)}$ correlation cannot be completely removed from the $g^{(2)}$ correlation. Moreover, the $g^{(1)}$ term is a fundamental ingredient for $g^{(2)}$ value by definition [23]. Recently, the physical origin of anticorrelation has been studied using the $g^{(1)}$ correlation via coherence rather than coincidence, emphasizing the wave nature of photons [24]. Based on this interpretation, the generation of nonclassical features using independent light sources can be easily understood via the $g^{(1)}$ correlation modified by an etalon, where the etalon creates $g^{(1)}$ dependence among independent photons within a narrow bandwidth [21]. In that sense, the $g^{(1)}$ correlation such as in Young's double-slit setup or a Mach-Zehnder interferometer (MZI) plays a key role in determining whether independent bipartite photons can be entangled or not via a specific phase relation between them [24].

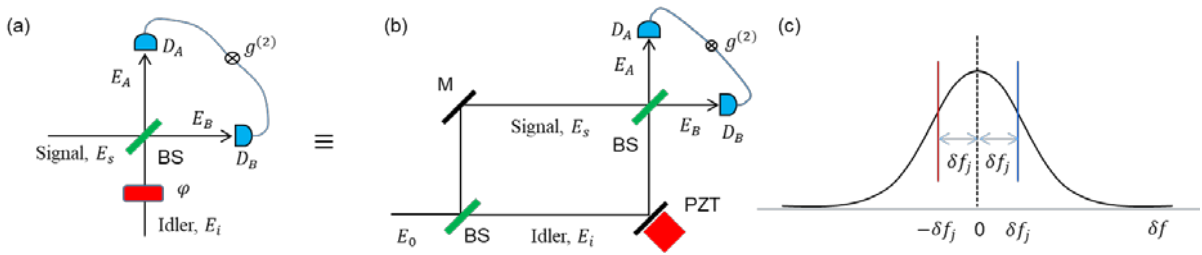

Fig. S1. Schematics of anticorrelation on a BS. (a) original scheme. (b) Modified one. (c) Spectral distribution of entangled photon pairs from SPDC processes. The subscript j stands for $j^{th}$ entangled photon pair. BS: nonpolarizing beam splitter. $D_k$: single photon detector for $E_k$.

Figure 1 shows a schematic of the typical anticorrelation on a BS, where two input photons are usually pre-entangled through spontaneous parametric down conversion (SPDC) processes. Figure 1(a) is the original scheme of, the so-called a HOM dip for $g^{(2)}(\tau) < 0.5$. Figure 1(b) is an equivalent scheme of Fig. 1(a) in MZI interferometry, where the specific phase relation between the two input photons can be achieved by a piezo-electric transducer (PZT). Figure 1(c) shows the bandwidth of the input photons for the signal and idler, where symmetric detuning $\pm \delta f_j$ across the line center occurs due to the energy conservation law between entangled photon pairs [2]. The line center indicates a half of the summed frequencies of each signal-idler photon pairs: $f_0 = f_s + f_i$, where $f_0$, $f_s$, and $f_i$ are the frequency of the pump, signal, and idler photons in SPDC, respectively. Thus, the following detuning relationship is always satisfied for all entangled photon pairs: $f_s^j = f_0 \pm \delta f_j$ and $f_i^j = f_0 \mp \delta f_j$. As a result, the summed error of the photon pair is dependent on the pump laser linewidth without violation of the uncertainty principle [25]. Here, the detuning sign of $\pm$ in $\delta f_j$ (denoted by red and blue lines) is random within the bandwidth. Considering this, the symmetrically detuned entangled photon pairs play a critical role in $g^{(2)}$ calculations, resulting in perfect cancellation of the $\lambda-$dependent $g^{(1)}$ term, where $\lambda$ is the carrier wavelength (discussed in the HOM analysis).

Regarding the SPDC process, there is a $\frac{\pi}{2}$ phase difference between the signal and idler photons according to the density matrix equations of quantum optics [26,27]. In a simple three-level atomic system composed of a ground, intermediate, and excited states, an excited atom experiences a $\pi-$phase shift with respect to the ground one. In the decay process, the atom gains a $\pi-$phase shift, resulting in no phase change when it competes a Rabi cycle. This relation is the physical background for the $\frac{\pi}{2}$ phase shift between the signal and idler photons. In nonlinear quantum optics, such a $\frac{\pi}{2}-$phase shift between the excited and intermediate atoms has already been demonstrated in [27-29]. In that sense, the idler photon $E_i$ in Fig. 1 can be denoted by $E_i = e^{i\pi/2}E_0$ in phasor notation, where $E_0$ represents the signal photon. For simplicity, the subscript is omitted in $E_0$ for different $f_j$ owing to the same $\frac{\pi}{2}-$phase shift. Considering another $\frac{\pi}{2}$ phase shift induced by a BS, each interacting entangled photon pair on a BS results in a destructive interference, where the $\pm$ sign of detuning in Fig. 1(c) randomizes the output direction of the bunched photons: $\langle I_A \rangle = \langle I_B \rangle$. As is well known, there is no $g^{(1)}-$caused interference in $g^{(2)}(\tau)$ measurements, where $g^{(1)}$ stands for the carrier wavelength ($\lambda$) dependency. Thus, the first goal here is to determine what makes $g^{(1)}$ disappear in the HOM dip.

Regarding coincidence detection, the time delay $\tau$ between $E_A$ and $E_B$ in Fig. 1 is actually between $E_s$ and $E_i$ because of $E_A = \frac{E_0}{\sqrt{2}}(1 + ie^{i\varphi})$ and $E_B = \frac{iE_0}{\sqrt{2}}(1 - ie^{i\varphi})$, where the phase $\varphi$ is used to control the idler photons [24]. For $\varphi = \frac{\pi}{2}$ and $\tau = 0$, $E_A = 0$ and $E_B = i\sqrt{2}E_0$. For $\varphi = -\frac{\pi}{2}$, the results are swapped. As the phase change occurs within the coincidence ($\tau = 0$) between $E_s$ and $E_i$ via a relative path-length control using a phase shifter or PZT, a $g^{(1)}-$caused interference fringe should be created in the outputs due to $0 \leq \varphi \leq 2\pi$: $I_A = \frac{I_0}{2}(1 - \cos(\varphi))$; $I_B = \frac{I_0}{2}(1 - \cos(\varphi))$. In all experimental demonstrations for the $g^{(2)}$ correlation, however, no such $\lambda-$depnedent interference fringe has been observed. Understanding this contradiction is key to the fundamental principles of anticorrelation as well as entanglement.

*Analysis of anticorrelation on a BS using photon pairs from SPDC*

Over the last several decades, $g^{(2)}$ correlation has been studied using SPDC- [2,30-32] and other method- [21,22] generated entangled photon pairs, where the amplitudes of the signal and idler photons are denoted by $E_s$ and $E_i$, repsectively: $E_s = E_0$ and $E_i = iE_0 e^{i\frac{2\pi}{c}\left[\frac{(f_0+\psi_j)}{2}+2\delta f_j\right]x}$. Although the pump frequency $f_0$ (denoted by '0' in the $\delta f$ axis) moves around back and forth due to its intrinsic linewidth, the $\delta f_j$ symmetry is intact. In Fig. 1(a), the idler photon is $\varphi-$phase controlled by a path-length change, resulting in the delay time $\tau$. This one side phase control is actually the same as the simultaneous control for both photons by BS [2]. Using the matrix representations, the following relation is obtained:

$$\begin{bmatrix} E_A \\ E_B \end{bmatrix} = \frac{1}{\sqrt{2}} e^{-i\varphi(\tau)/2} \begin{bmatrix} 1 & i \\ i & 1 \end{bmatrix} \begin{bmatrix} 1 & 0 \\ 0 & e^{i\varphi(\tau)} \end{bmatrix} \begin{bmatrix} E_s \\ E_i \end{bmatrix}, \qquad (1)$$

where $\varphi(\tau) = \frac{2\pi}{c}\left[\frac{(f_0+\psi_j)}{2} + 2\delta f_j\right]x$, and $x$ is the $\tau$−corresponindg path-length change by PZT. Here, $\psi_j$ is the initial phase given to the $j^{th}$ photon pair in SPDC processes. Because $\psi_j$ is random due to the spontaneous emission process, the sum of all interacting photon pairs for a certain period of time, e.g., one second results in $\varphi(\tau) \rightarrow \pm 2\delta f_j \tau$. Thus, $\varphi(\tau)$ is simply dependent upon $\delta f_j$, resulting in $I_A = I_0(1 \mp \cos\varphi)$ and $I_B = I_0(1 \pm \cos\varphi)$, where $I_0 = E_0 E_0^*$. This is the physical origin of why there is no $g^{(1)}$ correlation in most HOM dip measurements. Due to the equal chance of positive and negative signs in $\varphi(\tau)$ in the output intensity, the average value of each output intensity is $\langle I_A \rangle = \langle I_A \rangle = I_0$. Thus, the intensity correlation becomes:

$$g^{(2)}(\tau) = 2\langle \sin(\varphi)^2 \rangle, \qquad (2)$$

Because of $t_c \Delta = 1$ according to the uncertainty principle, the modulation depth of $\langle \sin(\varphi)^2 \rangle$ repidly decreases as $\tau\Delta$ increases, resulting in the transient effect as observed experimentally [32]. On the contrary of conventional understanding, this is a very important result in the interpretation of the HOM dip for the origin of $g^{(2)}(0)$ in terms of the wave nature.

Figure 2 shows numerical simulations of $g^{(2)}(\tau)$ using equation (2) for Fig. 1, where the FWHM of $g^{(2)}(\tau)$ is the same as that in Fig. 1(c). Based on the definition of intensity correlation $g^{(2)}(\tau)$, the maximum correlation occurs at $\tau = 0$. As $\tau$ increases, the average effect of the detuned pairs becomes effective, resulting in coherence washout equivalent to a lesser probability in photon bunching. Beyond the coherence time, $\tau > t_c$, the appeared wiggles in $g^{(2)}(\tau)$ are not an artifact but due to the averaged quantum beats between the signal and idler photons in $\delta f_j$. The disappearance of the wavelength $\lambda$−dependent interference fringe is due to the random phase given to each pair in SPDC processes. This is the correct answer to the question why there is no $\lambda$−dependent interference fringe in any of the HOM dip experiments.

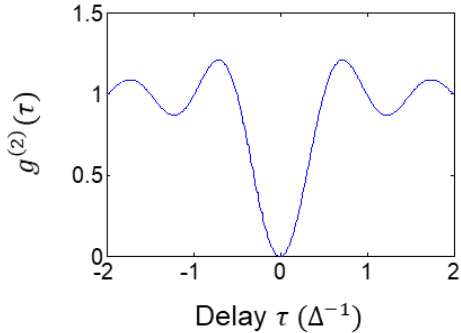

Fig. 2. Numerical simulation for intensity correlation for a HOM dip. Red dotted curve: Gaussian curve. $t_c$: coherence time. Arrows: FWHM. b=0 and c=15.

### *Analysis of $g^{(2)}$ correlation in an MZI using an attenuated laser*

Figure 1(b) as an equivalent scheme of Fig. 1(a) is now considered for the on-demand generation of anticorrelation, where any input field $E_0$ turns out to be $\frac{\pi}{2}$−phase shifted fields $E_s$ and $E_i$ via the BS, regardless of the input photon's frequency or jitter. To satisfy the coincidence detection, however, the input photon $E_0$ in Fig. 1(b) must be doubly bunched, where such bunched photons can be probabilistically obtained from an attenuated laser via Poisson statistics of the photon distribution. Then, the same scheme of Fig. 1(a) is achieved in Fig. 1(b), where the second BS results in interference between the two coherent photons for $g^{(2)}(\tau)$, where $\tau$ is controlled by PZT.

The MZI physics in Fig. 1(b) results in the following output photons, $E_A$ and $E_B$, where PZT-induced $\tau$ is proportional to the phase shift $\varphi$ $(\varphi_j = 2\pi f_j \tau)$ on $E_i$ with respect to $E_s$:

$$\begin{bmatrix} E_A \\ E_B \end{bmatrix} = \frac{1}{2}\begin{bmatrix} 1 - e^{i\varphi} & i(1 + e^{i\varphi}) \\ i(1 + e^{i\varphi}) & -(1 - e^{i\varphi}) \end{bmatrix}\begin{bmatrix} E_0 \\ 0 \end{bmatrix}, \tag{3}$$

where the subscript $j$ is omitted within the coherence time $t_c$ for simplicity. Thus, $E_A = \frac{E_0}{2}(1 - e^{i\varphi})$ and $E_B = \frac{iE_0}{2}(1 + e^{i\varphi})$ are resulted. The corresponding intensities are as follows:

$$I_A = I_0(1 - \cos(\varphi)), \tag{4}$$

$$I_B = I_0(1 + \cos(\varphi)). \tag{5}$$

Unlike the SPDC cases, the $\varphi$ in equations (4) and (5) is for the same frequency $(f_j)$ of the coherent photons without symmetric detuning $\pm \delta f_j$. For each $j^{th}$ photon, the initially given phase $\psi_j$ is the same each other. As is well known, the attenuated laser light is coherent, and thus the bunched photons are nearly perfectly identical in terms of frequency, phase, and jitter within the coherence time $t_c$.

Given the bandwidth of the coherent laser, the doubly bunched photons also have the same bandwidth. For commercially available lasers whose linewidth is $\Delta f = 10^{6\sim 12}$ Hz, the corresponding coherence time (length) is $\tau_c = 10^{-6}\sim 1$ μs ($l_c = 3\text{x}10^{-4}\sim 3\text{x}10^2$ m). Considering the wavelength of each photon in a visible spectrum, we can control each photon collectively within the coherence length $l_c$. Thus, the mean value of equations (4) and (5) should result in a $\lambda$−dependent oscillation without coherence washout in Fig. 2 (see Fig. 3). Unlike the symmetric detuning with a random initial phase in the SPDC case, the $g^{(1)}$ correlation term directly appears in $g^{(2)}(\tau)$. Thus, the expected anticorrelation should include a modulation fringe due to $g^{(1)}$, where the Gaussian envelope-caused coherence length is too long to be covered by the PZT scan range.

Regarding the normalized coincidence detection measurement rate $R_{AB}$ for the attenuated laser case, the following relation is obtained using equations (4) and (5):

$$R_{AB} = 1 - \cos(\varphi)^2, \tag{6}$$

where $R_{AB} = \frac{\langle I_A I_B \rangle}{\langle I_0 \rangle^2}$. In an attenuated laser governed by Poisson statistics, an antibunched photon stream can be easily obtained by adjusting the mean photon number to be $\langle n \rangle < 0.1$, where the ratio $\eta_{21}$ of doubly bunched photons to antibunched (single) photons is $\eta_{21} < 10^{-3}$. Keeping the number of single photons at $\sim 10^6$, then the number of doubly bunched photons is close to $\sim 10^3$, which is reasonable for $g^{(2)}$ measurements.

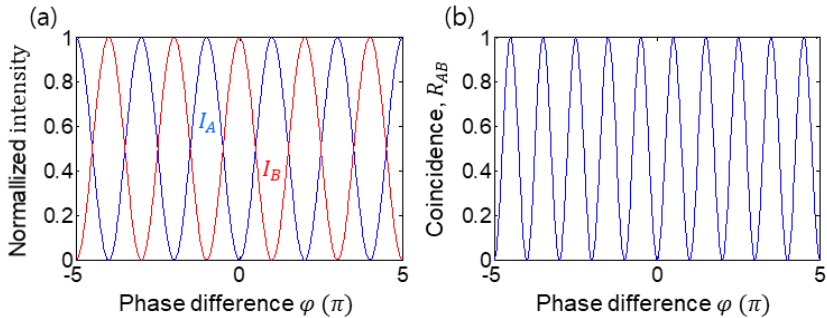

Fig. 3. Numerical simulations for (a) equations (4) and (5), and (b) equation (6).

Regarding the intensity correlation, $g^{(2)}(\tau) = \frac{\langle I_A I_B(\tau) \rangle}{\langle I_A \rangle \langle I_B(\tau) \rangle}$, however, both $\langle I_A \rangle$ and $\langle I_B \rangle$ in the denominator modulate according to equations (4) and (5), which has the same period as the numerator. Thus, the normalized coincidence detection $R_{AB}$ should be the same as $g^{(2)}$ if the undefined value of $\frac{0}{0}$ is avoided. This undefined case in $g^{(2)}$ calculations can be circumvented by considering a $\pi$−phase shifted $\varphi$, resulting in swapping between equations (4) and (5), without influence on the coincidence detection in equation (6). As a result, the following relation is satisfied:

$$g^{(2)}(\tau) = R_{AB}(\tau). \tag{7}$$

Figure 3 shows numerical simulations for the attenuated photons in Fig. 1(b). For this, the Gaussian envelop is neglected because the scan range is much shorter than the coherence length $l_c$. As shown in Fig. 3, the intensity correlation $g^{(2)}(\tau)$ modulates at a doubled rate compared with that in typical amplitude correlation $g^{(1)}$. Thus, the coincidence measurements include the $g^{(1)}$ term. This has never been discussed before and reveals fundamental understanding about the quantum features in anticorrelation for $g^{(2)}(0) = 0$. Here, the coincidence detection has no practical meaning without $g^{(1)}$ due to its long coherence time. This is because the wave nature plays an important role in the correlation measurements even in the HOM dip case. Because MZI does not distinguish between single photons and coherent light, a macroscopic quantum feature such as Schrödinger's cat can also be manipulated deterministically. Such an example has been discussed for photonic de Broglie waves using coherent light, recently [33]. Experimental demonstrations for Fig. 3 are discussed elsewhere [34].

*Discussion*

We have analyzed the fundamental difference between SPDC-generated photon pairs and attenuated coherent photons for the anticorrelation $g^{(2)}(\tau)$. In $g^{(2)}(\tau)$ of the SPDC-case, the symmetric detuning of the entangled photons (signal and idler) across the fixed half of the summed frequencies replaces the $g^{(1)}$ effect, where the original $g^{(1)}$ is disappeared due to the random phase given at the birth of entangled photon pairs. This is the physical origin of the disappeared $\lambda$−dependent $g^{(1)}$ in $g^{(2)}(\tau)$ in HOM dip measurements. In the attenuated laser case without the symmetric detuning, however, the $\lambda$−dependency should be vividly appeared on $g^{(2)}(\tau)$ due to coherence among all interacting photons. Because the coherence length of commercially available lasers is much longer than its wavelength $\lambda$, and photon indistinguishability for the quantum correlation $g^{(2)}(\tau)$ is sustained under coherence [35,36], the concept of coincidence detection in $g^{(2)}(\tau)$ is conceptually alleviated when the wave nature of photons is emphasized. In that sense, a collective control of entangled (wave) pairs can be achieved from a coherent system with a proper phase control [28,32,37].

*Conclusion*

We analyzed the conventional quantum feature of anticorrelation, the so-called HOM dip, based on SPDC-generated entangled photon pairs. With the symmetric detuning across the half summed frequencies of paired photons, the disappeared $g^{(1)}$ coherence was analyzed with intrinsic random phase given to each entangled photon pair due to the spontaneous emission process. Recently observed modulation ripples in both wings of the HOM dip [32] is the direct proof of the wave nature for a quantum beat as a special case of the symmetric detuning in SPDC processes. In attenuated coherent photons without symmetric detuning or random initial phase, the $\lambda$−dependent $g^{(1)}$ was appeared in $g^{(2)}(\tau)$ correlation. Based on the basic requirement of indistinguishability for quantum features, the $\lambda$−dependent modulation fringe in $g^{(2)}(\tau)$ indicates on-demand control of quantum correlation. This on-demand control has potential for the macroscopic quantum information processing. This research provides the fundamental physics of what entanglement should be and how to generate it.

Acknowledgments

This work was supported by GIST via GRI 2020.